\documentclass[Conference,a4paper]{IEEEtran}
\IEEEoverridecommandlockouts
\usepackage{color}
\usepackage{graphicx}
\usepackage{epstopdf}
\usepackage{amsmath}
\usepackage{amssymb}
\usepackage{algorithm}
\usepackage{algorithmic}
\usepackage{amsmath}
\usepackage{multirow}
\usepackage{booktabs}
\usepackage{array}
\usepackage{amsthm}
\usepackage{stfloats}
\usepackage{caption}
\usepackage{subfigure}
\usepackage{bm}
\usepackage{booktabs}
\usepackage{setspace}
\usepackage{diagbox}
\usepackage{enumerate}
\usepackage{ulem}
\usepackage{url}
\usepackage{circledsteps}
\usepackage{hyperref}
\allowdisplaybreaks[4]

{\bgroup
 \addtolength\abovedisplayshortskip{#1}
 \addtolength\abovedisplayskip{#1}
 \addtolength\belowdisplayshortskip{#1}
 \addtolength\belowdisplayskip{#1}
 }
{\egroup\ignorespacesafterend}
\newcommand{\be}{\begin{equation}}
\newcommand{\ee}{\end{equation}}
\newcommand{\bea}{\begin{eqnarray}}
\newcommand{\eea}{\end{eqnarray}}
\newcommand{\ba}{\begin{array}}
\newcommand{\ea}{\end{array}}

\title{
Resource Allocation for Transmissive RIS Transceiver Enabled SWIPT Systems
}
\author{\IEEEauthorblockN{Yuan Guo, Wen Chen, Xudong Bai, Chong He, and Qiong Wu
}
\thanks{
Y. Guo and W. Chen are with Department of Electronic Engineering, Shanghai Jiao Tong University, Shanghai, China, 
email:
yuanguo26@sjtu.edu.cn,
wenchen@sjtu.edu.cn. %}\thanks{
X. Bai is with the College of Microelectronics, 
Northwestern Polytechnical University, Xi'an, China, 
email: baixudong@nwpu.edu.cn. 
C. He is with the Department of Electronic Engineering, Shanghai Jiao Tong University, Shanghai, China, 
email: hechong@sjtu.edu.cn. 
Q. Wu is with the School of Internet of Things Engineering, Jiangnan University, Wuxi, China, 
emali: qiongwu@jiangnan.edu.cn.
}
}

\begin{document}
\maketitle
\pagestyle{empty}
\thispagestyle{empty}

\begin{abstract}
A novel transmissive reconfigurable intelligent surface (TRIS) transceiver-empowered simultaneous wireless information and power transfer (SWIPT) framework is proposed.
The sum-rate of the information decoding (ID) users is maximized by optimizing the TRIS transceiver's beamforming, 
subject to the energy harvesting (EH) users' quality-of-harvest 
and the per-antenna power constraints.
To solve this non-convex problem, 
we develop an efficient optimization algorithm. 
First, 
the original problem is reformulated as a semi-definite programming (SDP) problem. 
The resulting SDP problem
is then addressed using successive convex approximation (SCA) combined with a penalty-based method. 
Numerical results demonstrate the effectiveness of the algorithm.

\end{abstract}

\begin{IEEEkeywords}
Transmissive reconfigurable intelligent surface (TRIS) transceiver,
simultaneous wireless information and power transfer (SWIPT),
sum-rate maximization.
\end{IEEEkeywords}

\maketitle
%\vspace{-0.25em}
\section{Introduction}

With the explosive proliferation of Internet-of-Things (IoT) devices, 
simultaneous wireless information and power transfer (SWIPT) has emerged as a highly promising approach, 
drawing significant interest from both industry and academia \cite{ref_SWIPT_2}.
Acting as both information carriers and energy sources, 
radio-frequency (RF) signals in SWIPT systems enable concurrent data transmission and power delivery to IoT devices.

In recent years, 
a practical transmissive reconfigurable intelligent surface (TRIS) transceiver architecture \cite{ref_TRIS_1},
which consisting of a passive TRIS with a single horn antenna feed,
has been proposed as a cost- and energy-efficient enhancement for next-generation wireless networks.
In contrast to conventional multi-antenna base station (BS) that use many active RF elements, 
the TRIS configuration reduces hardware demands by eliminating multiple RF chains and complex signal processing modules.

Motivated by the TRIS transceiver concept,
a large body of research has explored leveraging TRIS technology to elevate the performance of wireless networks, 
e.g., \cite{ref_TRIS_app_1}$-$\cite{ref_TRIS_app_11}.
The work \cite{ref_TRIS_app_1} adopted a TRIS transceiver to enhance a multi-stream downlink communication system
and developed a low-complexity method to solve the max-min signal-to-interference-and-noise-ratio (SINR) problem.
In \cite{ref_TRIS_app_2},
the authors investigated 
a SWIPT system enabled by a TRIS transceiver,
which incorporates a nonlinear EH model and imperfect channel state information (CSI),
and addressed a sum-rate maximization problem.
The paper \cite{ref_TRIS_app_3} demonstrates that
integrating TRIS transceivers into the multi-tier computing network
can boost computing performance, decrease computation latency, and reduce BS deployment costs.
As reported in \cite{ref_TRIS_app_4},
a combined hybrid active-passive TRIS transceiver configuration was explored to achieve higher energy efficiency (EE) across the network.
The authors in \cite{ref_TRIS_app_5} investigated a cooperative TRIS-and-RIS strategy under CSI errors,
aiming to enhance the weighted sum secrecy capacity of the protected communication network.
In \cite{ref_TRIS_app_7}, 
the authors developed a TRIS-transceiver-based distributed cooperative ISAC solution 
aimed at extending the effective coverage of sensing and communication transceiver.
The work \cite{ref_TRIS_app_8} employed a TRIS transceiver design 
to enhance sum-rate performance in multi-cluster Low Earth Orbit (LEO) satellite systems using non-orthogonal multiple access
(NOMA) technique.
The TRIS transceiver-enabled multi-group multi-cast and multi-cell systems were researched in \cite{ref_TRIS_app_10}
and
\cite{ref_TRIS_app_11}, respectively.

Although the introduction of the TRIS transceiver brings significant performance improvements to wireless systems,
a TRIS transceiver-enabled SWIPT framework in which energy harvesting and information decoding are handled by separate receivers has not yet been investigated.
Motivated by this observation, 
we investigate the active beamforming design for a TRIS transceiver-enabled SWIPT system consisting of multiple ID and EH users.
In this work,
the contributions of this paper are summarized as follows:
\begin{itemize}
\item
This paper studies a novel TRIS transceiver-enabled SWIPT system 
in which the TRIS transceiver simultaneously transmits information to multiple ID users and supplies power to multiple EH users.
The objective is to maximize the sum-rate by optimizing the TRIS active beamforming.

\item
Due to the non-convexity of both the objective function and constraint,
the original problem is firstly transformed into a semi-definite programming (SDP) problem.
Then,
both the successive convex approximation (SCA) \cite{ref_MM}
and the penalty-based \cite{ref_Penalty} methods 
are invoked to solve non-convex objective function and rank-one constraints.

\item
Last but not least,
extensive simulation results are presented to validate the proposed algorithm's effectiveness 
in maximizing the SWIPT system's sum-rate. 
The results also show that the number of TRIS elements has a marked effect on system performance: 
increasing the element count yields higher sum-rate.

\end{itemize}

\section{System Model and Problem Formulation}
\subsection{System Model}

\begin{figure}[t]
	\centering
	\includegraphics[width=.30\textwidth]{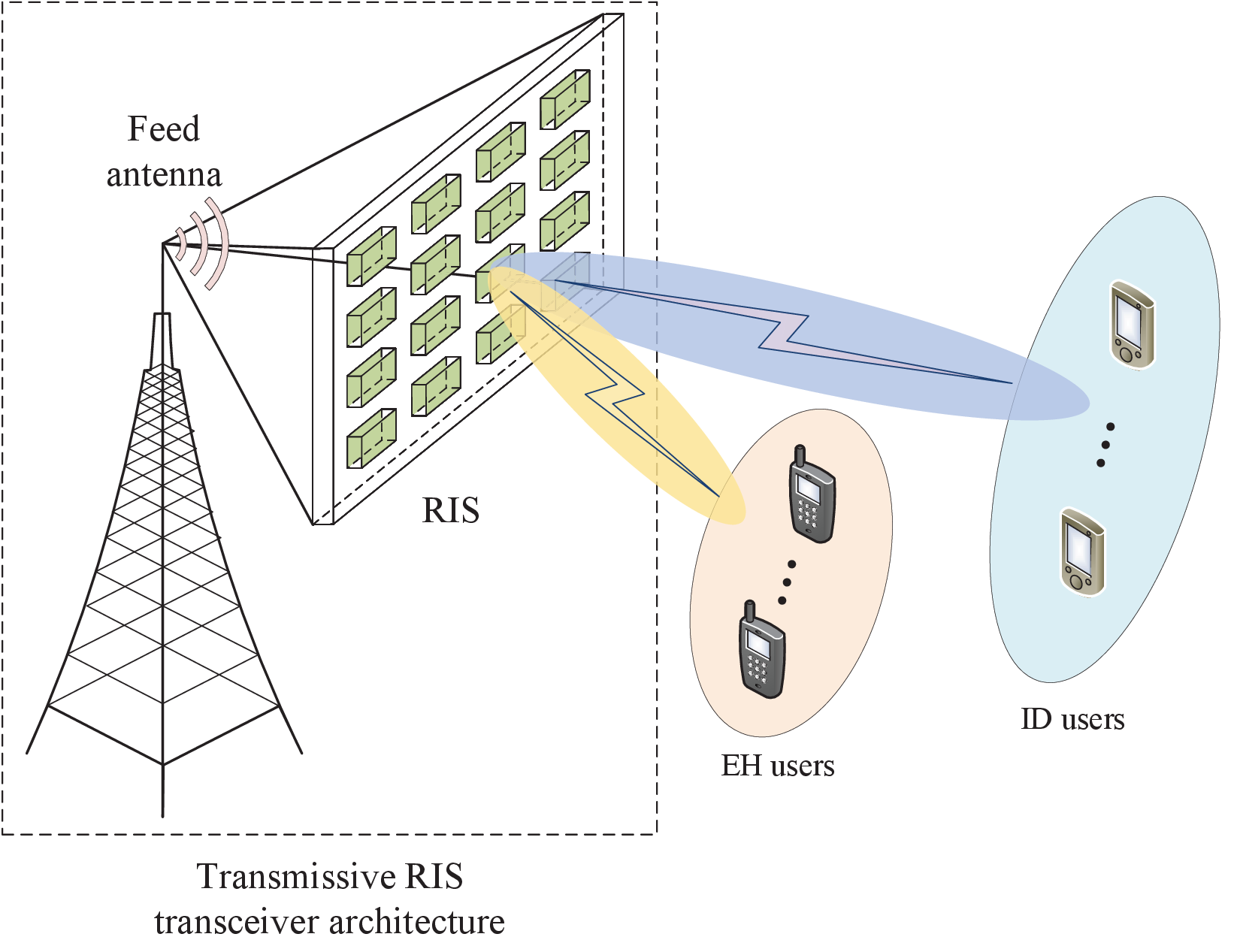}
	\caption{A TRIS transceiver-enabled SWIPT system.}
	\label{fig.1}
\end{figure}

A TRIS transceiver-enabled SWIPT system shown in Fig. \ref{fig.1} is investigated, 
where the TRIS transceiver, having $N$ antennas, 
is used to serve $K$ single-antenna ID users and $G$ single-antenna EH users simultaneously.
For simplicity,
the sets of ID users, EH users, and TRIS transceiver elements are respectively formulated as
$\mathcal{K} \triangleq \{1,\cdots,K\}$,
$\mathcal{G} \triangleq \{1,\cdots,G\}$
and
$\mathcal{N} \triangleq \{1,\cdots,N\}$.

The transmitted signal by the TRIS transceiver is written as 
\begin{align}
\mathbf{x} = {\sum}_{k=1}^{K} \mathbf{f}_{I,k}x_{I,k}
+
{\sum}_{g=1}^{G} \mathbf{f}_{E,g}x_{E,g},
\end{align}
where 
the information signal 
$x_{I,k} \in \mathbb{C}$ 
and energy signal
$x_{E,g} \in \mathbb{C}$
both follow the complex Gaussian distribution $\mathcal{CN}(0,1)$,
and the vectors 
$\mathbf{f}_{I,k}\in \mathbb{C}^{N\times 1}$
and 
$\mathbf{f}_{E,g}\in \mathbb{C}^{N\times 1}$
denote the beamforming for the $k$-th ID user and the $g$-th EH user, 
respectively.

Let 
$\mathbf{f}_{I}\triangleq [\mathbf{f}_{I,1}^T,\cdots,\mathbf{f}_{I,K}^T]^T \in \mathbb{C}^{NK \times 1}  $
and
$\mathbf{f}_{E}\triangleq [\mathbf{f}_{E,1}^T,\cdots,\mathbf{f}_{E,G}^T]^T \in \mathbb{C}^{NG \times 1}  $.
Under the signal generation way of the TRIS transceiver \cite{ref_TRIS_1},
the beamforming vectors $\mathbf{f}_{I}$ and $\mathbf{f}_{E}$ 
are constrained by per-antenna power limits, 
which can be written as
\begin{align}
\mathbf{f}_{I}^H \bar{\mathbf{A}}_{I,n} \mathbf{f}_{I} 
+ \mathbf{f}_{E}^H \bar{\mathbf{A}}_{E,n}\mathbf{f}_{E} \leq P_t, \forall n \in \mathcal{N},
\end{align}
where
$P_t$ denotes the maximum transmit power of the TRIS transceiver element.
The matrices $\bar{\mathbf{A}}_{I,n}$ and  $\bar{\mathbf{A}}_{E,n}$ can be respectively formulated as
\begin{align}
&\bar{\mathbf{A}}_{I,n} \triangleq \text{blkdiag}(\mathbf{A}_n, \cdots,\mathbf{A}_n  ) \in \mathbb{R}^{NK\times NK},\\
&\bar{\mathbf{A}}_{E,n} \triangleq \text{blkdiag}(\mathbf{A}_n, \cdots,\mathbf{A}_n  ) \in \mathbb{R}^{NG\times NG},
\end{align}
where
$\mathbf{A}_n \triangleq \text{diag}(\mathbf{a}_{n}) $
and the index vector $\mathbf{a}_{n}$ is denoted as
\begin{align}
\mathbf{a}_{n} \triangleq [0,0,\cdots,\underbrace{1}\limits_{\textrm{n-}th},\cdots,0,0] \in \mathbb{R}^{N\times N},
\end{align}
where the element at the $n$-th position is $1$ and all other elements are $0$.

In the following,
the received signal at the $k$-th ID user can be given as
\begin{align}
{y}_{I,k}
&=
\underbrace{{\mathbf{h}}_{I,k}^H\mathbf{f}_{I,k}x_{I,k}}
\limits_{\textrm{Desired signal}}
+\underbrace{{\sum}_{i\neq k}^{K}{\mathbf{h}}_{I,k}^H\mathbf{f}_{I,i}x_{I,i}}
\limits_{\textrm{Other ID users' interference}} \\
&+ \underbrace{{\sum}_{g=1}^{G} {\mathbf{h}}_{I,k}^H\mathbf{f}_{E,g}x_{E,g}}
\limits_{\textrm{EH users' interference}} 
 + n_{I,k},\nonumber
\end{align}
where 
$n_{I,k}  $ denotes the complex additive white Gaussian noise (AWGN) of the $k$-th ID user
and is distributed as $n_{I,k} \sim \mathcal{CN}(0, \sigma_{I,k}^2)$.
Let 
$ \bar{\mathbf{h}}_{I1,k} \triangleq [ \mathbf{h}_{I,k}^T,\cdots, \mathbf{h}_{I,k}^T ]^T \in \mathbb{C}^{NK \times 1}  $
and
$ \bar{\mathbf{h}}_{I2,k} \triangleq [ \mathbf{h}_{I,k}^T,\cdots, \mathbf{h}_{I,k}^T ]^T \in \mathbb{C}^{NG \times 1}  $.
Then the signal ${y}_{I,k}$ is rewritten as
\begin{align}
{y}_{I,k}
&=
\underbrace{\bar{\mathbf{h}}_{I,k}^H\mathbf{B}_{I,k}\mathbf{f}_{I}x_{I,k}}
\limits_{\textrm{Desired signal}}
+\underbrace{{\sum}_{i\neq k}^{K}\bar{\mathbf{h}}_{I,k}^H\mathbf{B}_{I,i}\mathbf{f}_{I}x_{I,i}}
\limits_{\textrm{Other ID users' interference}} \\
&+ \underbrace{{\sum}_{g=1}^{G} \bar{\mathbf{h}}_{I2,k}^H\mathbf{B}_{E,g}\mathbf{f}_{E}x_{E,g}}
\limits_{\textrm{EH users' interference}} 
 + n_{I,k},\nonumber
\end{align}
where 
$\mathbf{B}_{I,k} \triangleq \text{diag}(\mathbf{b}_{I,k}) \in \mathbb{R}^{NK\times NK}$
and
$\mathbf{B}_{E,g} \triangleq \text{diag}(\mathbf{b}_{E,g}) \in \mathbb{R}^{NG\times NG}$
are the selection matrices,
and the vector
$\mathbf{b}_{I,k}$ 
is defined as
\begin{align}
&\mathbf{b}_{I,k} \triangleq [0,\cdots,0,\underbrace{1,\cdots,1}
\limits_{N}
,0,\cdots,0]^T \in \mathbb{R}^{NK\times 1},
\end{align}
where entries from index 
$((k-1)\times N +1)$ to $ (k\times N )$ 
are 1 and every other component is 0. 
The vector $\mathbf{b}_{E,g}$  is constructed identically.

The SINR at the $k$-th ID user is represented as
\begin{align}
&\text{SINR}_{k}( \mathbf{f}_{I}, \mathbf{f}_{E}  )\\
&=
\frac{\vert\bar{\mathbf{h}}_{I,k}^H\mathbf{B}_{I,k}\mathbf{f}_{I}  \vert^2}
{
{\sum}_{i\neq k}^{K}\vert \bar{\mathbf{h}}_{I,k}^H\mathbf{B}_{I,i}\mathbf{f}_{I}\vert^2
+
{\sum}_{g=1}^{G} \vert \bar{\mathbf{h}}_{I2,k}^H\mathbf{B}_{E,g}\mathbf{f}_{E}\vert^2
+
\sigma_{I,k}^2
}.\nonumber
\end{align}
Hence,
the corresponding achievable rate of the $k$-th ID user is given by
\begin{align}
\mathrm{R}_{k}( \mathbf{f}_{I}, \mathbf{f}_{E}  ) =\text{log}\big( 1 + \text{SINR}_{k}( \mathbf{f}_{I}, \mathbf{f}_{E}  )\big).
\end{align}

We assume that EH users can harvest wireless energy from both the information and energy signals,
and adopt a linear EH model while neglecting the noise power \cite{ref_RIS_SWIPT_5}$-$\cite{ref_WMMSE_1}.
Thus,
the harvested energy at the $g$-th EH user is written as
\begin{align}
\mathrm{Q}_{g}( \mathbf{f}_{I}, \mathbf{f}_{E}  )
&=\zeta
\bigg(
\vert\bar{\mathbf{h}}_{E,g}^H\mathbf{B}_{E,g}\mathbf{f}_{E} \vert^2\\
&+
\underbrace{{\sum}_{i\neq g}^{G}
\vert\bar{\mathbf{h}}_{E,g}^H\mathbf{B}_{E,i}\mathbf{f}_{E} \vert^2}
\limits_{\textrm{Harvested energy from other EH signals}}
\nonumber\\
&+
\underbrace{
{\sum}_{k=1}^{K}\vert\bar{\mathbf{h}}_{E2,g}^H\mathbf{B}_{I,k}\mathbf{f}_{I} \vert^2}
\limits_{\textrm{Harvested energy from ID signals}}
\bigg),\nonumber
\end{align}
where
$\bar{\mathbf{h}}_{E,g} \triangleq [\mathbf{h}_{E,g}^T, \cdots ,\mathbf{h}_{E,g}^T  ]^T \in \mathbb{C}^{NG \times 1} $
and
$\bar{\mathbf{h}}_{E2,g} \triangleq [\mathbf{h}_{E,g}^T, \cdots ,\mathbf{h}_{E,g}^T  ]^T \in \mathbb{C}^{NK \times 1} $,
$0< \zeta \leq 1$ is the energy harvesting efficiency.

\subsection{Problem Formulation}

In this paper,
our objective is to maximize the sum-rate of all ID users by optimizing the active beamforming of the TRIS transceiver,
subject to a total harvested energy constraint of all EH users and 
the per-element transmit power constraints at the TRIS transceiver units.
Therefore,
the optimization problem is mathematically formulated as
\begin{subequations}
\begin{align}
\textrm{(P0)}:&\mathop{\textrm{max}}
\limits_{\mathbf{f}_I, \mathbf{f}_E
}\
{\sum}_{k=1}^{K}
\mathrm{R}_{k}(\mathbf{f}_I, \mathbf{f}_E) 
\label{P0_obj}\\
\textrm{s.t.}\ 
& {\sum}_{g=1}^{G}  \mathrm{Q}_g(\mathbf{f}_I, \mathbf{f}_E)  \geq Q_t,\label{P0_c_0}\\
& \mathbf{f}_{I}^H\mathbf{\bar{A}}_{I,n}\mathbf{f}_{I} + \mathbf{f}_{E}^H\mathbf{\bar{A}}_{E,n}\mathbf{f}_{E} \leq P_t, \forall n \in \mathcal{N},\label{P0_c_1}
\end{align}
\end{subequations}
where
(\ref{P0_c_0}) indicates the harvested energy constraint for all EH users 
with $Q_t$ denoting the minimum total harvested energy;
(\ref{P0_c_1}) is the transmit power constraint for each TRIS transceiver units.
The problem (P0) is difficult to solve since both the objective and the harvested-energy constraint are non-convex.

%\vspace{-0.2cm}
\section{Algorithm Design}

First,
to make the original problem (P0) more tractable,
we express the beamforming vectors as
\begin{align}
\mathbf{F}_{I} \triangleq \mathbf{f}_{I}\mathbf{f}_{I}^H \in \mathbb{C}^{NK\times NK},
\mathbf{F}_{E} \triangleq \mathbf{f}_{E}\mathbf{f}_{E}^H \in \mathbb{C}^{NG\times NG}.
\end{align}

Then,
the optimization problem (P1) can be equivalently reformulated as a rank-constrained SDP problem,
which is expressed as
\begin{subequations}
\begin{align}
\textrm{(P1)}:&\mathop{\textrm{max}}
\limits_{\mathbf{F}_I, \mathbf{F}_E
}\
{\sum}_{k=1}^{K}
\mathrm{\bar{R}}_{k}(\mathbf{F}_I, \mathbf{F}_E) 
\label{P1_obj}\\
\textrm{s.t.}\ 
& {\sum}_{g=1}^{G}  \mathrm{\bar{Q}}_g(\mathbf{F}_I, \mathbf{F}_E)  \geq Q_t,\label{P1_c_0}\\
& \text{Tr}( \mathbf{F}_{I}\mathbf{\bar{A}}_{I,n} ) + \text{Tr}( \mathbf{F}_{E}\mathbf{\bar{A}}_{E,n} ),\leq P_t, \forall n \in \mathcal{N},\label{P1_c_1}\\
&\mathbf{F}_{I} \succeq \mathbf{0},\ \mathbf{F}_{I}\succeq \mathbf{0},\label{P1_c_3}\\
& \text{Rank}(\mathbf{F}_{I}) = 1, \ \text{Rank}(\mathbf{F}_{E}) = 1, \label{P1_c_4}
\end{align}
\end{subequations}
where the objective function $\mathrm{\bar{R}}_{k}(\mathbf{F}_I, \mathbf{F}_E) $ and 
the function $\mathrm{\bar{Q}}_g(\mathbf{F}_I, \mathbf{F}_E)$ of constraint (\ref{P1_c_1}) 
can be rewritten in (\ref{P1_Obj_transformation_1}) and (\ref{P1_Obj_transformation_2}), respectively.
\begin{figure*}
%\begin{small}
\begin{align}
&\mathrm{\bar{R}}_{k}(\mathbf{F}_I, \mathbf{F}_E) \triangleq
\text{log}\bigg(1+
\frac{ \text{Tr}(\mathbf{B}_{I,k}\bar{\mathbf{h}}_{I,k}\bar{\mathbf{h}}_{I,k}^H\mathbf{B}_{I,k} \mathbf{F}_{I} ) }
{
{\sum}_{i\neq k}^{K}\text{Tr}(\mathbf{B}_{I,i}\bar{\mathbf{h}}_{I,k}\bar{\mathbf{h}}_{I,k}^H\mathbf{B}_{I,i} \mathbf{F}_{I} )
+
{\sum}_{g=1}^{G} \text{Tr}(\mathbf{B}_{E,g}\bar{\mathbf{h}}_{I2,k}  \bar{\mathbf{h}}_{I2,k}^H\mathbf{B}_{E,g}\mathbf{F}_{E}  )
+
\sigma_{I,k}^2
}
\bigg). \label{P1_Obj_transformation_1}
\end{align}
%\end{small}
\boldsymbol{\hrule}
\end{figure*}
\begin{figure*}
%\begin{small}
\begin{align}
\mathrm{\bar{Q}}_g(\mathbf{F}_I, \mathbf{F}_E) \triangleq
\zeta
\bigg(
{\sum}_{i=1}^{G}
\text{Tr}(\mathbf{B}_{E,i}\bar{\mathbf{h}}_{E,g} \bar{\mathbf{h}}_{E,g}^H\mathbf{B}_{E,i}\mathbf{F}_{E} )
+
{\sum}_{k=1}^{K}\text{Tr}(\mathbf{B}_{I,k}\bar{\mathbf{\mathbf{h}}}_{E2,g}\bar{\mathbf{h}}_{E2,g}^H\mathbf{B}_{I,k}\mathbf{F}_{I} )
\bigg).\label{P1_Obj_transformation_2}
\end{align}
%\end{small}
\boldsymbol{\hrule}
\end{figure*}

However,
as can be observed,
the difficulty in optimizing problem (P1)
stems from the objective function (\ref{P1_obj}) and the  rank-one constraints (\ref{P1_c_4}).
In the following,
we will address the objective function (\ref{P1_obj}) and the  rank-one constraints (\ref{P1_c_4}) one by one.

First, 
we reformulate the objective function (\ref{P1_obj}) as a difference-of-convex (DC) function, 
which can be expressed as follows
\begin{align}
\bar{\mathrm{R}}_{k}(\mathbf{F}_I, \mathbf{F}_E) = 
\dot{\mathrm{R}}_{k}(\mathbf{F}_I, \mathbf{F}_E) - \ddot{\mathrm{R}}_{k}(\mathbf{F}_I, \mathbf{F}_E),
\end{align}
where the functions 
$\dot{\mathrm{R}}_{k}(\mathbf{F}_I, \mathbf{F}_E)$ 
and
$\ddot{\mathrm{R}}_{k}(\mathbf{F}_I, \mathbf{F}_E)$
are respectively defined in (\ref{DC_form_1}) and (\ref{DC_form_2}).
\begin{figure*}
%\begin{small}
\begin{align}
&\dot{\mathrm{R}}_{k}(\mathbf{F}_I, \mathbf{F}_E)
\triangleq
\text{log}\bigg(
{
{\sum}_{i=1}^{K}\text{Tr}(\mathbf{B}_{I,i}\bar{\mathbf{h}}_{I,k}\bar{\mathbf{h}}_{I,k}^H\mathbf{B}_{I,i} \mathbf{F}_{I} )
+
{\sum}_{g=1}^{G} \text{Tr}(\mathbf{B}_{E,g}\bar{\mathbf{h}}_{I2,k}  \bar{\mathbf{h}}_{I2,k}^H\mathbf{B}_{E,g}\mathbf{F}_{E}  )
}
+
\sigma_{I,k}^2
\bigg)
, \label{DC_form_1}\\
&\ddot{\mathrm{R}}_{k}(\mathbf{F}_I, \mathbf{F}_E)
\triangleq
\text{log}\bigg(
{
{\sum}_{i\neq k}^{K}\text{Tr}(\mathbf{B}_{I,i}\bar{\mathbf{h}}_{I,k}\bar{\mathbf{h}}_{I,k}^H\mathbf{B}_{I,i} \mathbf{F}_{I} )
+
{\sum}_{g=1}^{G} \text{Tr}(\mathbf{B}_{E,g}\bar{\mathbf{h}}_{I2,k}  \bar{\mathbf{h}}_{I2,k}^H\mathbf{B}_{E,g}\mathbf{F}_{E}  )
}
+
\sigma_{I,k}^2
\bigg).\label{DC_form_2}
\end{align}
%\end{small}
\boldsymbol{\hrule}
\end{figure*}

Therefore,
the problem (P1) can be rewritten as
\begin{subequations}
\begin{align}
\textrm{(P2)}:&\mathop{\textrm{max}}
\limits_{\mathbf{F}_I, \mathbf{F}_E
}\
{\sum}_{k=1}^{K}
\bigg(
\dot{\mathrm{R}}_{k}(\mathbf{F}_I, \mathbf{F}_E) - \ddot{\mathrm{R}}_{k}(\mathbf{F}_I, \mathbf{F}_E)
\bigg) 
\label{P2_obj}\\
\textrm{s.t.}\ 
& ( \text{\ref{P1_c_0}})-(\text{\ref{P1_c_4}}).
\end{align}
\end{subequations}

Since the function 
$\ddot{\mathrm{R}}_{k}(\mathbf{F}_I, \mathbf{F}_E)$ 
is concave with respect to (w.r.t.) the variables 
$\mathbf{F}_I$
and
$\mathbf{F}_E$,
the objective function is non-convex.
Next,
according to the SCA method \cite{ref_MM},
we can obtain a tight upper-bounded by linearizing the non-convex function 
$\ddot{\mathrm{R}}_{k}(\mathbf{F}_I, \mathbf{F}_E)$ 
as follows
\begin{align}
&\mathrm{\ddot{R}}_k(\mathbf{F}_I, \mathbf{F}_E) \leq 
\mathrm{\ddot{R}}_k(\mathbf{F}_{I,0}, \mathbf{F}_{E,0}) \\
&+ \textrm{Tr}( \nabla_{\mathbf{F}_{I,0}}^H\mathrm{\ddot{R}}_k(\mathbf{F}_{I,0}, \mathbf{F}_{E,0})
(\mathbf{F}_{I}- \mathbf{F}_{I,0}) )\nonumber\\
&+ \textrm{Tr}( \nabla_{\mathbf{F}_{E,0}}^H\mathrm{\ddot{R}}_k(\mathbf{F}_{I,0}, \mathbf{F}_{E,0})
(\mathbf{F}_{E}- \mathbf{F}_{E,0}) ),\nonumber
\end{align}
where 
$\mathbf{F}_{I,0}$ 
and
$\mathbf{F}_{E,0}$ 
are obtained from the previous iteration,
and the gradients 
$\nabla_{\mathbf{F}_{I,0}}\mathrm{\ddot{R}}_k(\mathbf{F}_{I,0}, \mathbf{F}_{E,0})$
and
$\nabla_{\mathbf{F}_{E,0}}\mathrm{\ddot{R}}_k(\mathbf{F}_{I,0}, \mathbf{F}_{E,0})$ 
are respectively formulated in (\ref{Gradient_1}) and (\ref{Gradient_2}).
\begin{figure*}
%\begin{small}
\begin{align}
&\nabla_{\mathbf{F}_{I,0}}\mathrm{\ddot{R}}_k(\mathbf{F}_{I,0}, \mathbf{F}_{E,0})
\triangleq 
\frac
{
{\sum}_{i\neq k}^{K}\text{Tr}(\mathbf{B}_{I,i}\bar{\mathbf{h}}_{I,k}\bar{\mathbf{h}}_{I,k}^H\mathbf{B}_{I,i} )
}
{
{\sum}_{i\neq k}^{K}\text{Tr}(\mathbf{B}_{I,i}\bar{\mathbf{h}}_{I,k}\bar{\mathbf{h}}_{I,k}^H\mathbf{B}_{I,i} \mathbf{F}_{I} )
+
{\sum}_{g=1}^{G} \text{Tr}(\mathbf{B}_{E,g}\bar{\mathbf{h}}_{I2,k}  \bar{\mathbf{h}}_{I2,k}^H\mathbf{B}_{E,g}\mathbf{F}_{E}  )
+
\sigma_{I,k}^2
},
\label{Gradient_1}  \\
& \nabla_{\mathbf{F}_{E,0}}\mathrm{\ddot{R}}_k(\mathbf{F}_{I,0}, \mathbf{F}_{E,0})
\triangleq 
\frac
{
{\sum}_{g=1}^{G} \text{Tr}(\mathbf{B}_{E,g}\bar{\mathbf{h}}_{I2,k}  \bar{\mathbf{h}}_{I2,k}^H\mathbf{B}_{E,g}  )
}
{
{\sum}_{i\neq k}^{K}\text{Tr}(\mathbf{B}_{I,i}\bar{\mathbf{h}}_{I,k}\bar{\mathbf{h}}_{I,k}^H\mathbf{B}_{I,i} \mathbf{F}_{I} )
+
{\sum}_{g=1}^{G} \text{Tr}(\mathbf{B}_{E,g}\bar{\mathbf{h}}_{I2,k}  \bar{\mathbf{h}}_{I2,k}^H\mathbf{B}_{E,g}\mathbf{F}_{E}  )
+
\sigma_{I,k}^2
}\label{Gradient_2}.
\end{align}
%\end{small}
\boldsymbol{\hrule}
\end{figure*}

By applying the aforementioned transformations, 
the optimization problem (P2) can be expressed in a new form,
which can be given by
\begin{subequations}
\begin{align}
\textrm{(P3)}:&\mathop{\textrm{max}}
\limits_{\mathbf{F}_I, \mathbf{F}_E
}\
{\sum}_{k=1}^{K}
\bigg(
\dot{\mathrm{R}}_{k}(\mathbf{F}_I, \mathbf{F}_E) 
-
\mathrm{\ddot{R}}_k(\mathbf{F}_{I,0}, \mathbf{F}_{E,0}) \label{P3_obj}\\
&- \textrm{Tr}( \nabla_{\mathbf{F}_{I,0}}^H\mathrm{\ddot{R}}_k(\mathbf{F}_{I,0}, \mathbf{F}_{E,0})
(\mathbf{F}_{I}- \mathbf{F}_{I,0}) )\nonumber\\
&- \textrm{Tr}( \nabla_{\mathbf{F}_{E,0}}^H\mathrm{\ddot{R}}_k(\mathbf{F}_{I,0}, \mathbf{F}_{E,0})
(\mathbf{F}_{E}- \mathbf{F}_{E,0}) )
\bigg) \nonumber\\
\textrm{s.t.}\ 
& ( \text{\ref{P1_c_0}})-(\text{\ref{P1_c_4}}).
\end{align}
\end{subequations}

It is worth highlighting that the rank-one constraints (\ref{P1_c_4}) cause the optimization problem (P3) to be non-convex.
To deal with the rank-one constraints (\ref{P1_c_4}), 
we can respectively  rewrite the rank-one constraints 
$\text{Rank}(\mathbf{F}_{I}) = 1$
and
$ \text{Rank}(\mathbf{F}_{E}) = 1$
 as
follows
\begin{align}
&\text{Rank}(\mathbf{F}_{I}) = 1
\Leftrightarrow
\Vert\mathbf{F}_I\Vert_{\ast} - \Vert\mathbf{F}_I\Vert_{2} \leq 0,\\
&\text{Rank}(\mathbf{F}_{E}) = 1
\Leftrightarrow
\Vert\mathbf{F}_E\Vert_{\ast} - \Vert\mathbf{F}_E\Vert_{2} \leq 0,
\end{align}
where
$\Vert \cdot \Vert_{2}$
and
$ \Vert \cdot \Vert_{\ast} $
denote the spectral norm and nuclear norm, 
respectively.
Moreover,
considering the any positive semi-definite matrix,
the following inequalities can be achieved
\begin{align}
&\Vert\mathbf{F}_I\Vert_{\ast} = {\sum}_{i} \sigma_{I,i} \geq \Vert\mathbf{F}_I\Vert_{2}
 = \mathop{\text{max}}\limits_{i} \ \sigma_{I,i}, \label{Rank_one_transformation_1}\\
& \Vert\mathbf{F}_E\Vert_{\ast} = {\sum}_{i} \sigma_{E,i} \geq \Vert\mathbf{F}_E\Vert_{2}
 = \mathop{\text{max}}\limits_{i} \ \sigma_{E,i}, \label{Rank_one_transformation_2}
\end{align}
where
$\sigma_{I,i}$ and $\sigma_{E,i}$ denote the $i$-th singular values of the matrices $\mathbf{F}_I$ and $\mathbf{F}_E$, respectively.
Note that equality (\ref{Rank_one_transformation_1}) is valid when the rank of the matrix $\mathbf{F}_I$ equals one.
Similarly, 
the condition under which equality (\ref{Rank_one_transformation_2}) holds is analogous.

Next, 
by leveraging the penalty-based method \cite{ref_Penalty}, 
we convert the non-convex rank-one constraints
$\text{Rank}(\mathbf{F}_{I}) = 1$ and $\text{Rank}(\mathbf{F}_{E}) = 1$ 
into non-negative penalty terms that are appended to the objective function.
Consequently, 
problem (P3) can be transformed as follows
\begin{subequations}
\begin{align}
\textrm{(P4)}:&\mathop{\textrm{min}}
\limits_{\mathbf{F}_I, \mathbf{F}_E
}\
{\sum}_{k=1}^{K}
-\bigg(
\dot{\mathrm{R}}_{k}(\mathbf{F}_I, \mathbf{F}_E) 
-
\mathrm{\ddot{R}}_k(\mathbf{F}_{I,0}, \mathbf{F}_{E,0})\nonumber \\
&- \textrm{Tr}( \nabla_{\mathbf{F}_{I,0}}^H\mathrm{\ddot{R}}_k(\mathbf{F}_{I,0}, \mathbf{F}_{E,0})
(\mathbf{F}_{I}- \mathbf{F}_{I,0}) )\nonumber\\
&- \textrm{Tr}( \nabla_{\mathbf{F}_{E,0}}^H\mathrm{\ddot{R}}_k(\mathbf{F}_{I,0}, \mathbf{F}_{E,0})
(\mathbf{F}_{E}- \mathbf{F}_{E,0}) )
\bigg) \nonumber\\
&+\frac{1}{2\rho}
(
\Vert\mathbf{F}_I\Vert_{\ast}\! -\! \Vert\mathbf{F}_I\Vert_{2}
)   
+\frac{1}{2\rho}
(
\Vert\mathbf{F}_E\Vert_{\ast} \!-\! \Vert\mathbf{F}_E\Vert_{2}
) \label{P4_obj}  \\
\textrm{s.t.}\ 
& ( \text{\ref{P1_c_0}})-(\text{\ref{P1_c_3}}),
\end{align}
\end{subequations}
where $\rho > 0$ is the penalty factor.

Nevertheless, 
the non-convex nature of the penalty terms keep the reformulated problem (P4) challenging to solve.
Inspired by the SCA method,
we convexify the penalty terms 
by its first-order Taylor expansion to obtain the convex upper bounds,
which are given in (\ref{SCA_penalty_1}),
\begin{figure*}
%\begin{small}
\begin{align}
&\Vert\mathbf{F}_I\Vert_{\ast} - \Vert\mathbf{F}_I\Vert_{2} 
\leq
\Vert\mathbf{F}_I\Vert_{\ast}
-
\bigg\{
\Vert\mathbf{F}_{I,0}\Vert_{2}
+
\textrm{Tr}
\big( \boldsymbol{\lambda}_{max}(\mathbf{F}_{I,0})\boldsymbol{\lambda}_{max}^H(\mathbf{F}_{I,0}) 
(\mathbf{F}_{I} - \mathbf{F}_{I,0})  
\big)
\bigg\} 
%& 
\triangleq
\Vert\mathbf{F}_I\Vert_{\ast}
-
\bar{f}_I( \mathbf{F}_{I} ), 
%\nonumber
\label{SCA_penalty_1}\\
&\Vert\mathbf{F}_E\Vert_{\ast} - \Vert\mathbf{F}_E\Vert_{2} 
\leq
\Vert\mathbf{F}_E\Vert_{\ast}
-
\bigg\{
\Vert\mathbf{F}_{E,0}\Vert_{2}
+
\textrm{Tr}
\big( \boldsymbol{\lambda}_{max}(\mathbf{F}_{E,0})\boldsymbol{\lambda}_{max}^H(\mathbf{F}_{E,0}) 
(\mathbf{F}_{E} - \mathbf{F}_{E,0})  
\big)
\bigg\} 
%& 
\triangleq
\Vert\mathbf{F}_E\Vert_{\ast}
-
\bar{f}_E( \mathbf{F}_{E} )
, \nonumber
\end{align}
%\end{small}
\boldsymbol{\hrule}
\end{figure*}
where
$\boldsymbol{\lambda}_{max}(\mathbf{F}_{I,0})$
and
$\boldsymbol{\lambda}_{max}(\mathbf{F}_{E,0})$
denote the eigenvectors
corresponding to the largest eigenvalue of the matrices 
$\mathbf{F}_{I,0}$
and
$\mathbf{F}_{E,0}$,
respectively.

Therefore,
by replacing the penalty terms with the convex upper bounds presented in (\ref{SCA_penalty_1}),
the problem (P4) can be converted into the following optimization problem
\begin{subequations}
\begin{align}
\textrm{(P5)}:&\mathop{\textrm{min}}
\limits_{\mathbf{F}_I, \mathbf{F}_E
}\
{\sum}_{k=1}^{K}
-\bigg(
\dot{\mathrm{R}}_{k}(\mathbf{F}_I, \mathbf{F}_E) 
-
\mathrm{\ddot{R}}_k(\mathbf{F}_{I,0}, \mathbf{F}_{E,0})\nonumber \\
&- \textrm{Tr}( \nabla_{\mathbf{F}_{I,0}}^H\mathrm{\ddot{R}}_k(\mathbf{F}_{I,0}, \mathbf{F}_{E,0})
(\mathbf{F}_{I}- \mathbf{F}_{I,0}) )\nonumber\\
&- \textrm{Tr}( \nabla_{\mathbf{F}_{E,0}}^H\mathrm{\ddot{R}}_k(\mathbf{F}_{I,0}, \mathbf{F}_{E,0})
(\mathbf{F}_{E}- \mathbf{F}_{E,0}) )
\bigg) \nonumber\\
&+\frac{1}{2\rho}
\bigg(
\Vert\mathbf{F}_I\Vert_{\ast} - \bar{f}_I( \mathbf{F}_{I})
\bigg)   
+\frac{1}{2\rho}
\bigg(
\Vert\mathbf{F}_E\Vert_{\ast} - \bar{f}_E( \mathbf{F}_{E})
\bigg) \label{P5_obj}  \\
\textrm{s.t.}\ 
& ( \text{\ref{P1_c_0}})-(\text{\ref{P1_c_3}}).
\end{align}
\end{subequations}

Hence, the problem (P5) is convex and can be solved by existing convex optimization solvers, e.g., CVX \cite{ref_CVX}.
The proposed beamforming design algorithm can be summarized in Algorithm 1.

\begin{algorithm}[t]
\caption{The Penalty-based Method}
\label{alg:1}
\begin{algorithmic}[1]
\STATE {initialize}
$\mathbf{F}_{I}^{(0)}$,
$\mathbf{F}_{E}^{(0)}$
and
$t=0$
;
\REPEAT
\STATE update $\mathbf{F}_{I}^{(t+1)}$ and $\mathbf{F}_{E}^{(t+1)}$  by solving  (P5);
\STATE $t++$;
\UNTIL{$convergence$;}
\end{algorithmic}
\end{algorithm}

\section{Numerical Results}

\begin{figure}[t]
	\centering
	\includegraphics[width=.30\textwidth]{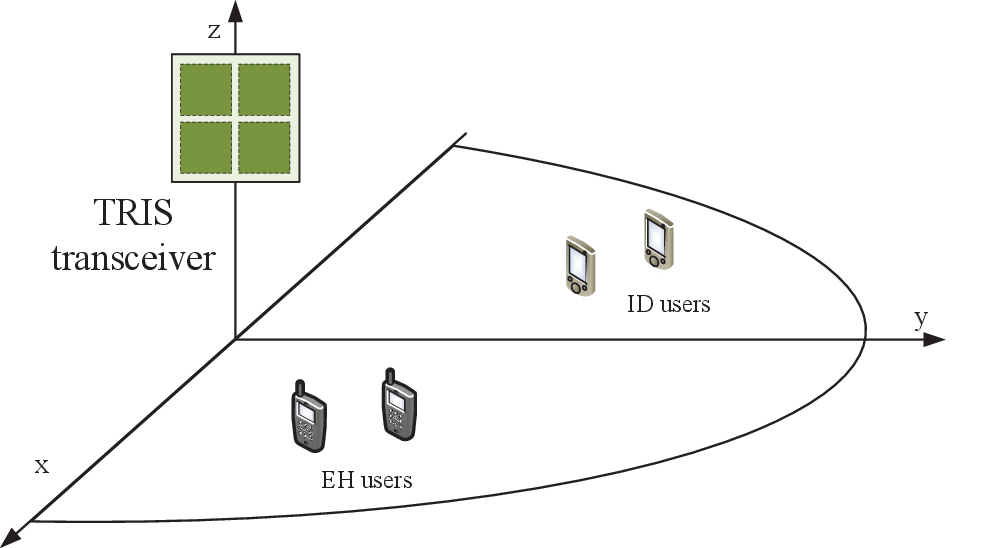}
	\caption{The simulation setup.}
	\label{fig.2}
\end{figure}

In this section, 
we evaluate the performance of the proposed beamforming design algorithm.
The simulated setup is illustrated in Fig. \ref{fig.2},
where a TRIS transceiver consisting of $N=36$ units serves $K=2$ ID users and $G=2$ EH users simultaneously.
We consider a 3D coordinate system in which the TRIS transceiver is located at (0, 0, 1.5)m, 
and all ID users are randomly distributed within a sector at radial distances between 20m and 50m and positioned at a height of 1.5m.
The pathloss exponent for the TRIS transceiver-ID user link is set to 3.2, 
whereas that for the TRIS transceiver-EH user link is 2.2.
The per-unit transmit power limit of the TRIS transceiver is 10dBm, 
and the noise power at each ID user is -90dBm.

\begin{figure}[t]
	\centering
	\includegraphics[width=.32\textwidth]{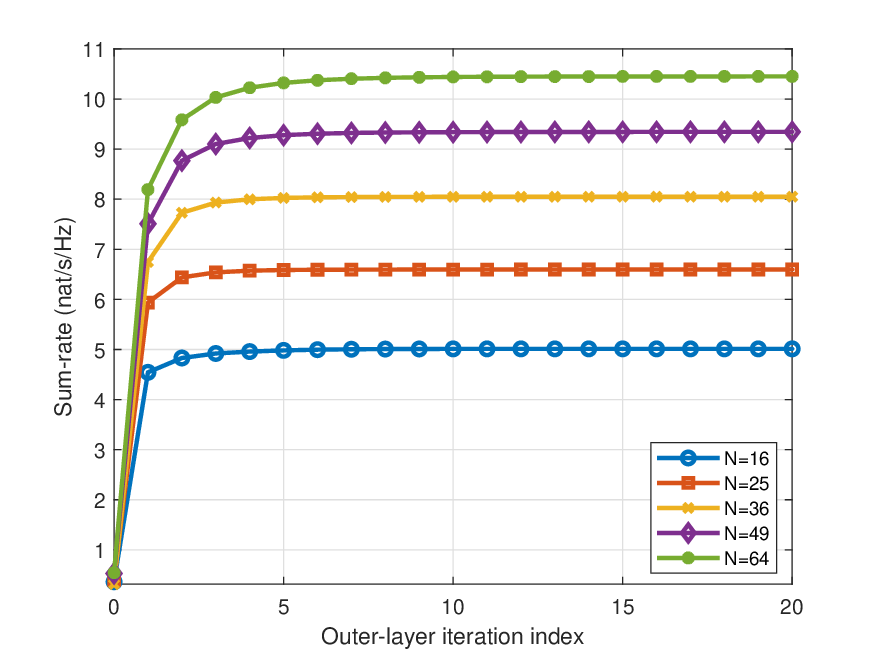}
	\caption{Convergence of Algorithm.}
	\label{fig.3}
\end{figure}

Fig. \ref{fig.3} shows the convergence performance of the Alg. \ref{alg:1} under various settings of the TRIS transceiver units.
As can be observed,
the sum-rate increases monotonically with the iteration index and generally converges within 10 iterations.

\begin{figure}[t]
	\centering
	\includegraphics[width=.32\textwidth]{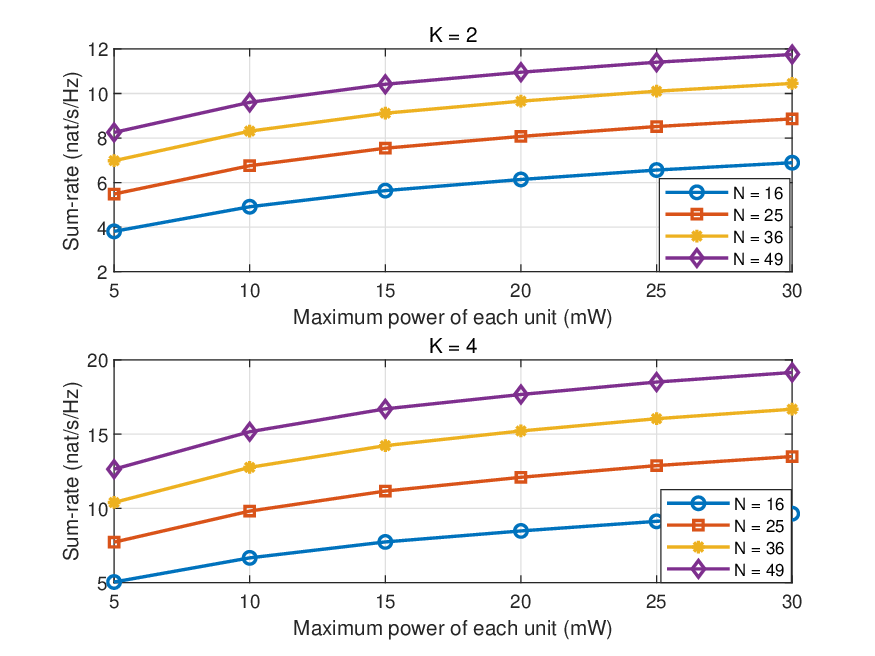}
	\caption{Impact of the maximum power of each TRIS element.}
	\label{fig.4}
\end{figure}

In Fig. \ref{fig.4}, 
we study the impact of the maximum power of TRIS unit on sum-rate.
The upper and lower subplots correspond to different numbers of ID users, respectively.
It is evident from Fig. \ref{fig.4} that the sum-rate grows monotonically as the TRIS unit's maximum transmit power increases.

\begin{figure}[t]
	\centering
	\includegraphics[width=.32\textwidth]{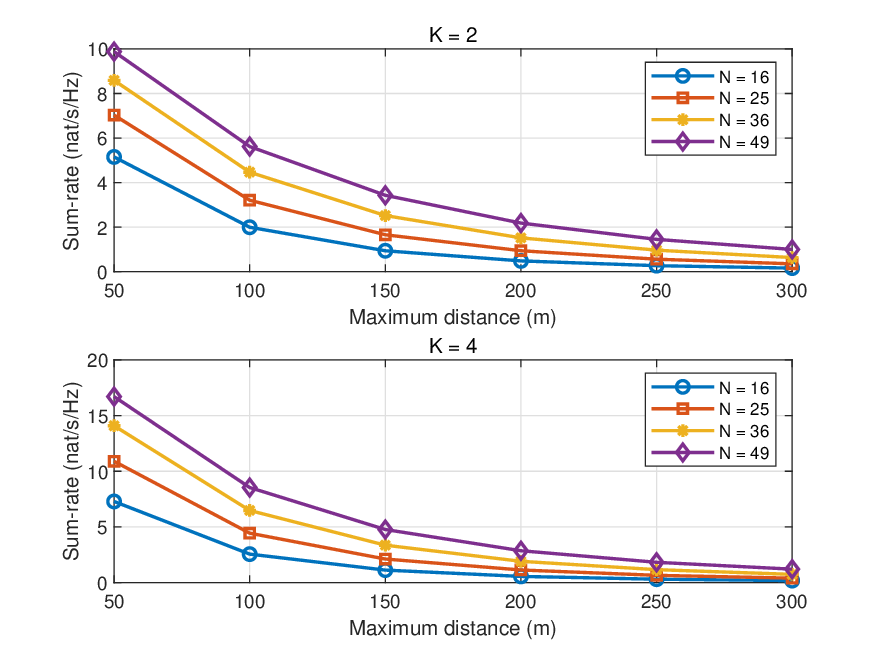}
	\caption{Impact of the maximum distance between TRIS and ID users.}
	\label{fig.5}
\end{figure}

In Fig. \ref{fig.5}, 
we compare the sum-rate versus the maximum distance between TRIS and ID users.
For all array sizes $N$, 
we observe that
the sum-rate decreases monotonically as the distance increases from 50m to 300m, 
with the steepest drop occurring between 50m and 100m.

\section{Conclusions}
%\vspace{0.2 cm}

In this article,
we studies the TRIS transceiver-aided SWIPT system.
To enhance the spectral efficiency,
a sum-rate maximization problem is formulated by optimizing the TRIS transceiver active beamforming.
In order to solve the non-convex optimization problem,
an efficient iterative algorithm is developed by leveraging the penalty-based and SCA methods.
Simulation results have verified the tremendous potential of the TRIS transceiver
to improve system performance of SWIPT systems.

%\appendix
%\subsection{Proof of Theorem 1}
\normalem
%

%\vspace{-3.85em}

%\enlargethispage{-6.5cm}
\end{document}